\documentclass[prd,amsmath,amssymb,nofootinbib]{revtex4} 
\usepackage{graphicx} 
\usepackage{amsmath}
\usepackage{amsfonts,amsbsy}
\usepackage{amssymb}

\def\vp#1{\mathbf{#1}}
\newcommand{\be}{\begin{equation}}
\newcommand{\ee}{\end{equation}}
\newcommand{\bea}{\begin{eqnarray}}
\newcommand{\eea}{\end{eqnarray}}

\renewcommand{\d}{\mathrm{d}}
\def\dd#1{\frac{\mathrm{d}^2\mathbf{#1}}{(2\pi)^2}}

\def\lsim{\mathrel{\rlap{\lower4pt\hbox{\hskip1pt$\sim$}}
    \raise1pt\hbox{$<$}}}                
\def\gsim{\mathrel{\rlap{\lower4pt\hbox{\hskip1pt$\sim$}}
    \raise1pt\hbox{$>$}}}                

\begin{document}
\title{The small-$x$ gluon distribution in centrality biased $pA$ and
  $pp$ collisions}

\author{Adrian Dumitru}
\affiliation{Department of Natural Sciences, Baruch College, CUNY,
17 Lexington Avenue, New York, NY 10010, USA}
\affiliation{The Graduate School and University Center, The City
  University of New York, 365 Fifth Avenue, New York, NY 10016, USA}
\affiliation{Physics Department, Brookhaven National
  Laboratory, Upton, NY 11973, USA}

\author{Gary Kapilevich}
\affiliation{The Graduate School and University Center, The City
  University of New York, 365 Fifth Avenue, New York, NY 10016, USA}

\author{Vladimir Skokov}
\affiliation{RIKEN-BNL Research Center, Brookhaven National
  Laboratory, Upton, NY 11973, USA}

\begin{abstract}
The nuclear modification factor $R_{pA}(p_T)$ provides information on
the small-$x$ gluon distribution of a nucleus at hadron
colliders. Several experiments have recently measured the nuclear
modification factor not only in minimum bias but also for central $pA$
collisions. In this paper we analyze the bias on the configurations of
soft gluon fields introduced by a centrality selection via the number
of hard particles. Such bias can be viewed as reweighting of
configurations of small-$x$ gluons.  We find that the biased nuclear
modification factor ${\cal Q}_{pA}(p_T)$ for central collisions is
above $R_{pA}(p_T)$ for minimum bias events, and that it may redevelop
a ``Cronin peak'' even at small $x$. The magnitude of the peak is
predicted to increase approximately like $1/{A_{\perp}}^\nu$,
$\nu\sim0.6\pm0.1$, if one is able to select more compact
configurations of the projectile proton where its gluons occupy a
smaller transverse area $A_\perp$. We predict an enhanced ${\cal
  Q}_{pp}(p_T)-1 \sim 1/(p_T^2)^\nu$ and a Cronin peak even for
central $pp$ collisions.
\end{abstract}

\maketitle

\section{Introduction}

The nuclear modification factor $R_{pA}(k)$ of the single-inclusive
transverse momentum distribution in proton-nucleus collisions has
received intense scrutiny from theory as well as experiment. It
provides insight into the small-$x$ gluon distributions of a nucleus
at high-energy hadron colliders. $R_{pA}(k)$ is defined as
\begin{equation}  \label{eq:RpA_def}
  R_{pA}(k) = \frac{\left<\frac{\d N}{\d^2 k \, \d y}\right>\Big|_{pA}}
  {N_\mathrm{coll}^\mathrm{m.b.} \, \left<\frac{\d N}{\d^2 k\,\d y}
    \right>\Big|_{pp}}~.
\end{equation}
Thus, it is given by the ratio of the single-inclusive transverse
momentum distributions in {\em minimum bias} $pA$ versus $pp$
collisions, scaled by the corresponding number of binary collisions
which is proportional to the average thickness $\sim A^{1/3}$ of the
target nucleus. In the absence of nuclear effects, $R_{pA}(k)=1$.

In the high-energy limit where particle production is dominated by
soft, small-$x$ gluons with semi-hard transverse momenta, $R_{pA}(k)$
is suppressed~\cite{RpA,RpA_AAKSW}. At high transverse momentum beyond the
saturation scale $Q_{s,A}$ of the nucleus this is leading twist shadowing due to
the fact that the gluon distribution acquires a BFKL anomalous
dimension~\cite{bfkl}, in the presence of a saturation
boundary~\cite{Mueller:2002zm}, which differs from its asymptotic
DGLAP limit~\cite{DGLAP}. We review the basic argument for how the
suppression arises in sec.~\ref{sec:pA-g-Xsection}.

A suppression of $R_{pA}(k)$ has been observed in the central region
of $p+Pb$ collisions at 5~TeV for hadron transverse momenta below
about 2~GeV~\cite{LHC-RpA,LHC-RpA-CMS,Aad:2016zif}. (Transverse
momenta degrade in gluon fragmentation by roughly a factor of 2.) The
data is described reasonably well by
models~\cite{LHC-RpA-rcBK1,LHC-RpA-rcBK} which employ unintegrated
gluon distributions which solve the running coupling BK
equations~\cite{rcBK}\footnote{Also see the compilation of predictions
  for the $p+$Pb 5~TeV LHC run published in ref.~\cite{Albacete:2013ei}
  and the follow-up comparison to data~\cite{Albacete:2016veq}.}. These
models predict a stronger suppression out to higher transverse
momentum at forward rapidities.  For d+Au collisions at 200~GeV, the
BRAHMS and STAR collaborations at RHIC have found a suppression of
$R_{dAu}(k)$ at forward rapidities~\cite{RHIC_RdAu}. Note that the
parton transverse momentum distribution is significantly steeper in
the forward region of collisions at RHIC energy than in the central
region at LHC energies. Consequently, the typical hadron momentum
fraction $\langle z\rangle$ in fragmentation is greater at lower
energy and higher rapidity.

One may also analyze the nuclear modification factor in {\em central}
$pA$ collisions. Naively, this corresponds to collisions where the
projectile proton suffers an inelastic collision with a greater than
average number of target nucleons. This would be analogous to minimum
bias $pA$ collisions with a target nucleus with many more than $\sim
200$ nucleons (which, of course, is not available). Accordingly, one
expects a {\em stronger suppression} of $R_{pA}(k)$ for central versus
minimum bias events. This qualitative expectation is confirmed by
model calculations~\cite{LHC-RpA-rcBK} for events with
$N_\mathrm{part,Pb}\ge 10$, which exceeds the average $\left<
N_\mathrm{part,Pb} \right> \approx7$ for minimum bias $p+$Pb collisions
at 5~TeV. However, neither the number of target participants nor the
number of collisions can be measured directly. Experimentally, one
therefore employs a variety of different centrality measures.

Several collaborations have analyzed the nuclear modification factor
for central collisions. Their main observation is that it {\em
  increases} with centrality and that it displays a Cronin like peak
at transverse momenta of $3\dots4$~GeV. This is in qualitative
disagreement with the naive expectation described above.  The
ATLAS collaboration~\cite{Aad:2016zif} classifies the centrality of
the events according to the transverse energy deposited in the
hemisphere of the nucleus at $-4.9<\eta<-3.1$. The number of collisions
$N_\mathrm{coll}$ in each centrality class is estimated from  Glauber
models with or without Glauber-Gribov corrections.

\begin{figure}[htb]
 \includegraphics[width=10cm]{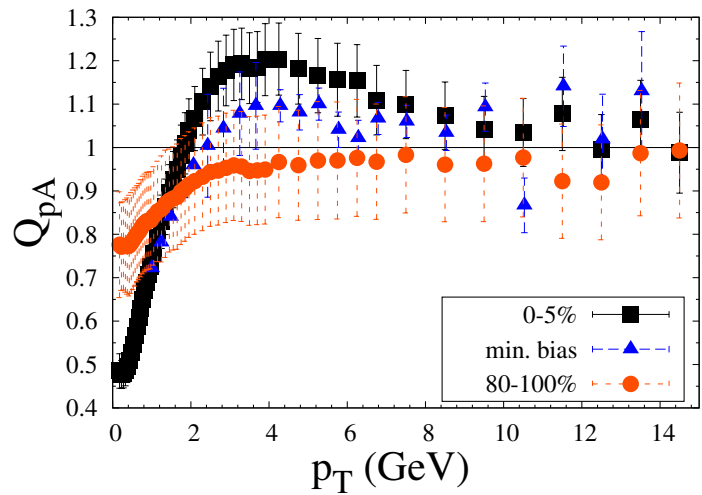}
 \caption{ALICE data~\cite{Adam:2014qja} for $Q_{pA}(p_T)$ in
   ``central'', minimum bias, and ``peripheral'' $p+$Pb collisions at
   5.02~TeV for $p_T<15$~GeV. For these data sets they define
   centrality classes with zero degree calorimeters, and determine the
   number of collisions $N^{\mathrm{mult}}_\mathrm{coll}$ in each
   centrality class from the charged particle multiplicity at
   midrapidity.}
 \label{fig:QpA_ALICE}
 \end{figure}
The ALICE collaboration employed several different methods to quantify
collision centrality~\cite{Adam:2014qja}. Their ``hybrid'' methods
define centrality classes based on the signal of a zero degree
calorimeter in the hemisphere of the nucleus which should classify
events according to the number of participant nucleons. The number of
binary collisions in a particular centrality class is then determined by either
one of the following models:
\begin{enumerate}
\item the charged-particle multiplicity at mid-rapidity is
  proportional to the number of participants from which one obtains
  $N_\mathrm{coll}^\mathrm{mult}$;
\item the yield of charged high-$p_T$ particles at mid-rapidity is
  proportional to the number of binary NN collisions ($\to
  N^\mathrm{high-pT}_\mathrm{coll}$);
\item the target-going charged-particle multiplicity is proportional
  to the number of wounded target nucleons from which one obtains
  $N^\mathrm{Pb-side}_\mathrm{coll}$.
\end{enumerate}
The ALICE collaboration denotes the nuclear modification factor in the
presence of a centrality selection by ${\cal Q}_{pA}$.
 
Our main point is that the experimental centrality selections
discussed above introduce a bias on the small-$x$ gluon configurations
producing such events. For example, $pA$ events with higher than
average multiplicity or transverse energy deposition not only involve
more nucleon participants in the target. They also involve a biased
average over the ensemble of small-$x$ gluon fields. This bias has to
be accounted for also in calculations of the transverse momentum
distributions. The biased average can be interpreted as {\em
  reweighting} of configurations,
\bea
\left< O\right> &=& \frac{\sum_i w_i O_i}{\sum_i w_i}~~~~,~~~~
w_i = e^{-S_i}~,\\
\to ~~
\left< O\right>_\mathrm{bias} &=& \frac{\sum_i w'_i O_i}{\sum_i w'_i}
~~~~,~~~~ w'_i = w_i b_i ~.  \label{eq:<O>_bias}
\eea
Here, $i=1\dots N$ labels field configurations, $O$ denotes an operator
corresponding to a particular observable\footnote{For example,
  eq.~(\ref{eq:spectrum1}) provides the operator which generates the
  single-inclusive gluon distribution.}, $S_i$ is the action
evaluated on configuration $i$, and $b_i$ is the configuration
bias. Note that the $O_i$ as well as the operator averages in general
are functions of transverse momentum, rapidity etc., and that the
functional form of $\left< O\right>(p_T)$ will differ from that of
$\left< O\right>_\mathrm{bias} (p_T)$ even when the reweighting
parameters $b_i$ are $p_T$-independent numbers. The predictive power
of the approach corresponds to precisely this modification of the
$p_T$-dependence of the expectation value of the observable due to
reweighting. The percentile of configurations obtained from the
reweighted ensemble is given by
\be
\nu_r = \frac{\left(\sum_i w'_i\right)^2}{N\sum_i (w'_i)^2}~.
\ee

While we are unable to precisely replicate the centrality
selections employed by the experiments, we illustrate our point within
the following setting. We focus on the observable
\begin{equation}  \label{eq:QpA_def}
  {\cal Q}_{pA}(k) = \frac{\left<\frac{\d N}{\d^2 k \, \d y}\right>_\mathrm{bias}\Big|_{pA}}
  {N_\mathrm{coll}^\mathrm{bias} \, \left<\frac{\d N}{\d^2 k\,\d y}
    \right>\Big|_{pp}}~.
\end{equation}
The transverse momentum distribution in the denominator is that for
minimum bias $pp$ collisions; that in the numerator corresponds to the
class of $pA$ (or $pp$) collisions for a particular bias. We choose the number
of hard gluons in the linear regime above the so-called ``extended
geometric scaling'' scale, $p_T>Q_{gs,A}\gg Q_{s,A}$, as our ``centrality''
selector. (Of course, such a centrality selector can not be used by
the experiments because the number of such hard particles per event is
small.)

Thus, configurations of the target corresponding to a
greater than average number of such high transverse momentum gluons
are selected as ``central'' collisions. Also, we define the number of
collisions for this biased sample to be given by the ratio of the
number of gluons above $Q_{gs,A}$ for a heavy-ion versus a proton
target:
\begin{equation}
  N_\mathrm{coll}^\mathrm{bias} = \frac{\int\limits_{Q_{gs,A}} \d^2 k
    \left<\frac{\d N}{\d^2 k \,
      \d y}\right>_\mathrm{bias}\Big|_{pA}}
  {\int\limits_{Q_{gs,A}} \d^2 k \left<\frac{\d N}{\d^2 k\,\d y}
    \right>\Big|_{pp}}~.
\end{equation}
This is motivated by one of ALICE's methods for determining the number
of collisions in a particular centrality class from the multiplicity
of high transverse momentum particles,
$N^{\mathrm{high-}p_T}_\mathrm{coll}$, and derives from the fact that
the number of hard particles in minimum-bias collisions is
proportional to $N_\mathrm{coll}$. However, we emphasize that here we
do not consider Glauber model fluctuations of $N_\mathrm{coll}$ (for
simplicity). Rather, here $N_\mathrm{coll}^\mathrm{bias} >
N_\mathrm{coll}^\mathrm{m.b.}$ arises entirely due to the bias on the
small-$x$ gluon configurations. Given this centrality selector our
goal is to analyze the transverse momentum distribution {\em below}
$Q_{gs,A}$ resulting from the biased average over soft gluon
fields. Our main result is that, indeed, ${\cal Q}_{pA}(k)$ for a high
multiplicity bias is greater than the minimum-bias $R_{pA}(k)$ at
small $x$.

The importance of ``color charge fluctuations'' in models of the
initial state for hydrodynamics of A+A collisions has been pointed out
in refs.~\cite{AA_color_flucs}. There, one is concerned mainly with
the spatial distribution of the $p_T$-integrated gluon (energy)
density which determines the pressure gradients.  Here, we focus
instead on the biased nuclear modification factor which provides
direct insight into the $p_T$ dependence of the gluon distribution
functions in a reweighted JIMWLK ensemble.

We should stress that even though the bias effect on the small-$x$
fields discussed here must certainly exist, that does not imply that
understanding the data would not require other, additional
effects. For example, it is possible that ``color reconnections''
(see~\cite{Sjostrand:2013cya} and references therein) become more
important as one selects more ``central'' events. Here, however, we
focus on illustrating the conceptual issue of biased averages over the
ensemble of small-$x$ gluon fields rather than to attempt a definitive
interpretation of the data. The specific question we ask is how a
multiplicity bias imposed at high transverse momentum affects the
distribution of small-$x$ gluons at lower $k$.

This paper is organized as follows. In the following section we
discuss basic features of biased averages over the ensemble of
small-$x$ fields, and we present qualitative analytic estimates of
${\cal Q}_{pA}$. Section~\ref{sec:JIMWLKnum} presents numerical
results obtained by Monte-Carlo sampling of the JIMWLK ensemble. The
final section is devoted to a brief summary and discussion.

\section{Cross section for gluon production} \label{sec:pA-g-Xsection}
To leading order in the gluon density of the projectile proton the
single inclusive gluon distribution is given by~\cite{CGC_pA}
\begin{align}
\left<  E_k \frac{d N}{d^3 k}\right>  &=
  \frac{1}{(2\pi)^3 k^2} 
  (
  \delta_{ij} \delta_{lm} +  \epsilon_{ij} \epsilon_{lm} 
  )
 \left< \ \Omega_{ij}^b(\vp{k}) 
  \ \left[\Omega_{lm}^b(\vp{k})\right]^* \right>_{p,A}
  \notag \\ &=
  \frac{g^2}{(2\pi)^3 k^2} 
  (
  \delta_{ij} \delta_{lm} +
  \epsilon_{ij} \epsilon_{lm} 
  ) 
  \int 
  {\dd p}
  {\dd q}
  \frac{p_{ i} (k-p)_{j} }{p^2}
  \frac{q_{l} (k-q)_{m} }{q^2}
  \notag \\ 
  &\quad\quad\quad \times
 \left< \rho_a^*(\vp{q}) 
  \left[
    W^\dagger(\vp{k}-\vp{q})
    W(\vp{k}-\vp{p}) 
    \right]_{ab}
  \rho_b(\vp{p})  \right>_{p,A}~.    \label{eq:spectrum1}
\end{align}
Here 
\begin{equation}
  \Omega_{ij}^a (\vp{x}) = g \left[ \frac{\partial_i}{\partial^2}
    \rho^b(\vp{x}) \right] \partial_j W^{ab}(\vp{x}) ~,
  \label{eq:Omega}
\end{equation}
and
\begin{equation}
W(\vp{x}) = {\cal P}\, e^{-ig\int dx^- A^{+a}(x^-,\vp{x})\, {T^a}}
\end{equation}
is a Wilson line in the adjoint representation in the field of the
target. The brackets in eq.~(\ref{eq:spectrum1}) denote an average
over the random color charge densities or soft gluon fields of both
projectile and target\footnote{$\rho^a$ in
  eqs.~(\ref{eq:spectrum1},\ref{eq:Omega}) denotes the color charge
  density of the proton projectile. We have traded that of the target
  in favor of its classical field $A^{+a}$ in covariant gauge.}.

For simplicity, we assume that the gluon distribution of the
projectile proton can be approximated by the most likely
distribution. It is straightforward, in principle, to allow for a bias
on the gluon distribution of the proton along the same lines as for
the nucleus.  Thus, we average over all configurations of the proton
projectile using
\begin{equation}
\left< \rho_a^*(\vp{q}) \, \rho_b(\vp{p})\right>_p = \delta_{ab}\, (2\pi)^2
\delta(\vp{q}-\vp{p})\, \mu_p^2(q^2)~.
\end{equation}
Then the gluon spectrum~(\ref{eq:spectrum1}) becomes
\begin{equation}
  \left< E_k \frac{d N}{d^3 k} \right> =
  \frac{g^2}{(2\pi)^3 k^2} 
  \int 
  {\dd{p}}
  \frac{(\vp{k}-\vp{p})^2 }{p^2}\, \mu^2_p(p^2)
 \left< \mathrm{tr} \,
    W^\dagger(\vp{k}-\vp{p})
    W(\vp{k}-\vp{p}) 
  \right>_A~.    \label{eq:spectrum2}
\end{equation}
For high transverse momentum, $k\gg Q_{s,A}$, we can evaluate this
integral in the leading logarithmic approximation:
\begin{eqnarray}
  \left<E \frac{dN}{d^3 k}\right>
  &=& \frac{1}{2 (2\pi)^3} \frac{1}{g^2C_F} \frac{Q_{s,p}^2}{k^4}
  \log \left(\frac{k^2}{Q_{s,p}^2}\right) \times
k^4  \left< \mathrm{tr} \,
  W^\dagger (\vp{k}) W(\vp{k})\right>_A ~.   \label{Eq:LOapp}
\end{eqnarray}
We have introduced the saturation momentum of the
proton via $Q_{s,p}^2 = C_F g^4\mu_p^2/2\pi$ which acts as an infrared
cutoff because the initial assumption that the proton be dilute does not
apply below $Q_{s,p}$. At high transverse momentum the last factor on
the r.h.s.\ of eq.~(\ref{Eq:LOapp}) is approximately constant.

For intermediate transverse momenta of order $Q_{s,A}$ on the other
hand we can simplify eq.~(\ref{eq:spectrum2}) to
\begin{eqnarray}
  \left<E \frac{dN}{d^3 k}\right>
  &=& \frac{1}{2 (2\pi)^3} \frac{1}{g^2C_F} \frac{Q_{s,p}^2}{k^2}
  \log \left(\frac{Q_{s,A}^2}{Q_{s,p}^2}\right) \times
   k^2  \left< \mathrm{tr} \,
  W^\dagger (\vp{k}) W(\vp{k})\right>_A ~.   \label{Eq:LOapp_k2}
\end{eqnarray}
Again, for $k\sim Q_{s,A}$ the last factor is approximately
constant. In what follows we will restrict to $k>Q_{s,A}$, however,
and work with eq.~(\ref{Eq:LOapp}).

It will be useful to recall first how the nuclear suppression of
\begin{equation}  \label{eq:RpA}
  R_{pA}(k) = \frac{\left<\frac{\d N}{\d^2 k\d y}\right>\Big|_{pA}}
  {N_\mathrm{coll}^\mathrm{m.b.} \, \left<\frac{\d N}{\d^2 k\d y}\right>\Big|_{pp}}
\end{equation}
arises in minimum bias collisions at high transverse momentum
$k>Q_{s,A}$ and small $x$. This is due to the fact that the gluon distribution
acquires an anomalous dimension at small $x$ which leads to ``leading
twist shadowing''. For detailed discussions we refer to
refs.~\cite{RpA,RpA_AAKSW}. At high transverse momentum we can expand the
correlator of Wilson lines in eq.~(\ref{Eq:LOapp}) to quadratic order
in $gA^+$ which leads to
\begin{equation}
k^4  \left< \mathrm{tr}\, W^\dagger(\vp{k}) W(\vp{k}) \right> \to
  g^4 N_c (N_c^2-1) \, \mu^2(k) A_\perp
~. \label{eq:avgW+W_leading}
\end{equation}
The transverse area $A_\perp$ in this expression is determined by the
projectile proton. Also,
\begin{equation} \label{eq:mu2_q2}
\mu^2(k) \simeq \mu_0^2 \left(\frac{k^2}{Q_{s}^2}\right)^{1-\gamma(k)}
\end{equation}
when $k>Q_{s}$~\cite{Iancu:2002aq}. $\gamma(Q_s)\equiv\gamma_s
\simeq0.64$ is the BFKL anomalous dimension~\cite{bfkl} in the
presence of a saturation boundary~\cite{Mueller:2002zm}; it increases
logarithmically with increasing $k/Q_s$ and is close to its asymptotic
DGLAP limit of 1 at $k^2 \sim Q_{gs}^2\sim
Q_s^4/\Lambda^2$~\cite{Levin:2000mv,Iancu:2002tr}\footnote{$Q_{gs}^2\sim
  Q_s^4$ is the asymptotic behavior at high rapidity. Attempts have
  been made to incorporate pre-asymptotic corrections in
  phenomenological parametrizations of
  $\gamma(k)$~\cite{Dumitru:2005kb}. However, in this section we do
  not present numerical predictions and therefore we leave aside such
  issues.}. Note that for a nucleus $\mu_0^2\sim A^{1/3}$ is
proportional to the average thickness and hence to
$N_\mathrm{coll}^\mathrm{m.b.}$. Then, using
eqs.~(\ref{Eq:LOapp},\ref{eq:avgW+W_leading}) in~(\ref{eq:RpA}) we get
\begin{equation}  \label{eq:RpA_mb}
  R_{pA}(k) = \frac{\mu^2_A(k)}{N_\mathrm{coll}^\mathrm{m.b.}\,
    \mu^2_p(k)}
  \simeq (k^2)^{\gamma_p(k)-\gamma_A(k)}\,
  \frac{(Q_{s,p}^2)^{1-\gamma_p(k)}}{(Q_{s,A}^2)^{1-\gamma_A(k)}}
  \simeq \left(\frac{k^2}{Q_{s,p}^2}\right)^{\gamma_p(k)-\gamma_A(k)}\,
  \frac{1}{(N_\mathrm{coll}^\mathrm{m.b.})^{1-\gamma_A(k)}}  ~.
\end{equation}
For a given $k$ the anomalous dimension for the gluon distribution of
the nucleus differs from that of the proton since $k/Q_s$ is not the
same. Hence, we distinguish $\gamma_p(k)$ from $\gamma_A(k)$.

The previous equation confirms the somewhat stronger suppression of
the {\em minimum bias} $R_{pA}$ for a thicker target mentioned in the
introduction. Also, since $\gamma_p(k)>\gamma_A(k)$ at midrapidity it
follows that $R_{pA}(k)$ increases with $k$ towards its asymptotic
value of $R_{pA}(k)=1$ where both anomalous dimensions are close to
1. We emphasize that eqs.~(\ref{eq:mu2_q2},\ref{eq:RpA_mb}) refer to
the fixed point of the small-$x$ renormalization group, i.e.\ to high
rapidity where memory of the initial condition has been lost.

Eqs.~(\ref{Eq:LOapp},\ref{Eq:LOapp_k2}) show that the ensemble of
gluon spectra in a $pA$ collision is determined by the ensemble of
adjoint dipole correlators. Rather than computing the unbiased
inclusive cross section averaged over all target configurations we are
interested here in performing that average subject to a {\em
  constraint} on the covariant gauge gluon distribution, i.e.\ that
$g^2 \mathrm{tr}\, A^+(\vp q) A^+(-\vp q)$ be equal to a given
function $X(\vp q)$. The distribution of functions $X(\vp q)$ in the
JIMWLK ensemble is described by the constraint effective
potential~\cite{Dumitru:2017cwt}
\begin{eqnarray}
  e^{-V_\mathrm{eff}[X(\vp q)]} &=& \int {\cal D}A^+(\vp q) \,
  W_Y[A^+(\vp q)] \,
\delta(X(\vp q)-g^2 \mathrm{tr}\, A^+(\vp q) \, A^+(-\vp q))~.
\end{eqnarray}
We can evaluate $V_\mathrm{eff}[X(\vp q)]$ analytically in a Gaussian
approximation where~\cite{Iancu:2002aq}
\begin{eqnarray}
  W_Y[A^+(\vp q)] = e^{-S_G[A^+(\vp q)]}~~~~~,~~~~~
  S_G[A^+(\vp q)] &=& \int {\dd q} \, q^4\, \frac{\mathrm{tr}\, A^+(\vp q)\,
A^+(-\vp q)}{g^2 \mu^2(q^2)}~.
\end{eqnarray}
$\mu^2(q^2)$ evolves with rapidity but we shall leave this dependence
implicit.  With the Gaussian action for $A^+$ one obtains (see
ref.~\cite{Dumitru:2017cwt} for details)
\begin{eqnarray}
V_\mathrm{eff}[X(\vp q)] &=& \int \dd q
\left[\frac{q^4}{g^4\mu^2(q)}X(\vp q)
  - \frac{1}{2} A_\perp N_c^2\log X(\vp q)\right]~.  \label{eq:Veff_A+A+}
\end{eqnarray}
The extremal configuration is determined by $\delta
V_\mathrm{eff}[X(q)] / \delta X(\ell)=0$ to be
\begin{equation} \label{eq:Xs}
  X_s(\ell) = \frac{1}{2} N_c^2 A_\perp \frac{g^4\mu_A^2(\ell)}{\ell^4}~.
\end{equation}
This is equal to $\left< X(\ell)\right>$ up to terms of order ${N_c}^0$.
The partition sum is
\begin{equation}
  Z = \int {\cal D}X(\vp q)\, e^{-V_\mathrm{eff}[X(\vp q)]}~,
\end{equation}
and the unbiased expectation value of an observable $O[g^2 \mathrm{tr}\,
  |A^+(\vp q)|^2]$ can be written as
\be  \label{eq:<O>_DX}
\left< O\right> = \frac{1}{Z} 
\int {\cal D}X(\vp q)\, e^{-V_\mathrm{eff}[X(\vp q)]}\, O[X(\vp q)]~.
\ee
Below, integrals over $X(\vp q)$ are understood to be normalized by
$1/Z$ as appropriate. A {\em biased} average of $O$ can be obtained by
restricting the set of functions $X(\vp q)$ that one integrates over in
eq.~(\ref{eq:<O>_DX}). More generally, one may reweight like in
eq.~(\ref{eq:<O>_bias}) by letting $\exp(-V_\mathrm{eff} [X(\vp q)])
\to \exp(-V_\mathrm{eff} [X(\vp q)]) \, b[X(\vp q)]$. For
\be
b[X] = \exp\left(
\int \d^2 {\vp q}\, t(\vp q) X(\vp q)
\right)~,
\ee
in particular, $Z[t]=\left<1\right>_t$ is a generating functional for the
correlation functions of $X(\vp q)$:
\be
\frac{1}{Z} \frac{\delta^n Z[t]}{\delta t(\vp k_1)\cdots \delta t(\vp
  k_n)}\Bigr|_{t\equiv 0} =
\left< X(\vp k_1)\cdots X(\vp k_n)\right>~.
\ee

\subsection{The distribution of functions
  $\mathrm{tr} \,  W^\dagger (\vp{k}) W(\vp{k})$ at order $(gA^+)^2$}
\label{sec:WW_LO}
  
We now determine the distribution of the functions   $\mathrm{tr} \,
W^\dagger (\vp{k}) W(\vp{k})$. To do so, we expand the Wilson lines in
powers of the covariant gauge field $gA^+$. To order $(gA^+)^2$,
\begin{eqnarray}
  \left< \mathrm{tr}\, W^\dagger(\vp{k}-\vp{p}) W(\vp{k}-\vp{p})
  \right> &=&
-\frac{g^4 N_c (N_c^2-1) A_\perp}{(2\pi)^2} \int \d^2\vp r \,
e^{-i(\vp{k}-\vp{p})\cdot\vp r}
\int\frac{\mathrm{d}^2\vp s}{s^4} \mu^2_A(s) \left(1-e^{i\vp s\cdot\vp
  r}\right) \nonumber\\
& & \times
  \int {\cal D}X(\vp q) \, e^{-V_\mathrm{eff}[X(q)]} \, \frac{X(\vp
    s)}{X_s(s)} \label{eq:O(A+^2)}\\
  && \hspace{-2cm}=
  g^4 N_c (N_c^2-1) A_\perp \int\frac{\mathrm{d}^2\vp s}{s^4}
  \left[\delta(\vp k - \vp p
    -\vp s) - \delta(\vp k - \vp p)\right]\, \mu^2_A(s)
  \int {\cal D}X(\vp q) \, e^{-V_\mathrm{eff}[X(q)]} \, \frac{X(\vp s)}{X_s(s)}~.
\end{eqnarray}
By inspection of eq.~(\ref{eq:spectrum2}) the ``tadpole'' subtraction $\sim \delta(\vp k
- \vp p)$ does not contribute and can be dropped:
\begin{equation}
  \left< \mathrm{tr}\, W^\dagger(\vp{k}-\vp{p}) W(\vp{k}-\vp{p})
  \right> \to
  \frac{g^4 N_c (N_c^2-1)}{(\vp k - \vp p)^4} \, \mu^2_A(|\vp k - \vp p|) A_\perp
  \int {\cal D}X(\vp q) \,
  e^{-V_\mathrm{eff}[X(q)]} \,
  \frac{X(\vp k - \vp p)}{X_s(|\vp k - \vp p|)}~. \label{eq:W+W_leading}
\end{equation}
Also, if the distribution $e^{-V_\mathrm{eff}[X(q)]}$ of functions
$X(\vp q)$ is assumed to be infinitely strongly peaked about $X_s(q)$
then the integral over $X(\vp q)$ is equal to 1 (recall that we
suppress the $Z^{-1}$ normalization factor for ease of notation), and
we recover eq.~(\ref{eq:avgW+W_leading}) from above.

Since we expanded the product of Wilson lines to leading non-trivial
power of $A^+$ only, we must use eq.~(\ref{eq:W+W_leading}) for
the high-$k$ gluon spectrum. Hence, eq.~(\ref{Eq:LOapp}) now reads
\begin{eqnarray}
  \left<E \frac{dN}{d^3 k}\right>_{\mathrm{high-}k}
  &=& \frac{g^2N_c^2  \mu^2_A(k) A_\perp}{(2\pi)^3}  \frac{Q_{s,p}^2}{k^4}
  \log \left(\frac{k^2}{Q_{s,p}^2}\right)
  \int {\cal D}X(\vp q) \,
  e^{-V_\mathrm{eff}[X(q)]} \,
  \frac{X(\vp k)}{X_s(k)} ~.   \label{Eq:LOapp_Xa}
\end{eqnarray}
To understand qualitatively the effect due to the bias towards
functions $X(\vp q)$ corresponding to a greater than average number of
hard gluons we assume that the saddle point of the reweighted (biased)
ensemble is shifted to
\begin{eqnarray}
  \frac{X(\vp q)}{X_s(q)} &=& 1 + \eta_0 \left(\frac{{q_0}^2}{q^2}\right)^a\,
  \Theta(Q^2-q^2)\,
  \Theta(q^2-Q^2_{s,A})~,~~~~~~~~~(a,\eta_0 \ge0).
  \label{eq:trialX}
\end{eqnarray}
Such a shift could be obtained by reweighting with
\be
\log b[X] \simeq \frac{1}{2}A_\perp N_c^2 \, \eta_0
  \int\limits_{Q_\mathrm{gs}^2}^{Q^2} \dd{\ell}
\frac{X(\ell)-X_s(\ell)}{X_s(\ell)} \left(\frac{{q_0}^2}{\ell^2}\right)^a
  + {\cal O}\left({\eta_0}^2\right)~.
\ee
While this functional reweights configurations based only on the
number of gluons above the extended geometric scaling scale
$Q_\mathrm{gs}$, we shall see in sec.~\ref{sec:JIMWLKnum} that, in
fact, a spectral shape like in eq.~(\ref{eq:trialX}) emerges
automatically all the way down to scales of order $Q_{s,A}$.

Our trial function~(\ref{eq:trialX}) increases the number of gluons in
the transverse momentum range between $Q_{s,A}$ and $Q$. The parameter
$\eta_0$ corresponds to the amplitude of the enhancement, and $Q\sim
Q_{gs,A}$ determines how far in transverse momentum it extends. $q_0$
is some fixed transverse momentum scale, for instance
$Q_{s,p}(Y=0)$. Obviously, ${q_0}^{2a}$ could be absorbed into
$\eta_0$ but writing $X(\vp q)$ in the form~(\ref{eq:trialX}) appears
clearer.

Such $X(\vp q)$ comes with a penalty action of
\begin{equation}
V_\mathrm{eff}[X(q)] \simeq \frac{1}{8\pi} N_c^2 A_\perp \, q_0^2\, \eta_0
\times
\begin{cases}
  \frac{1}{1-a} \left(\frac{Q^2}{q_0^2}\right)^{1-a} & (0\le a<1)~, \\
  \log\frac{Q^2}{Q_{s,A}^2} & (a=1)~.
\end{cases}
\end{equation}
Without reweighting it is suppressed by a relative probability
\begin{equation}
p_r =   \frac{w[X(\vp q)]}{w[X_s(q)]} = e^{-V_\mathrm{eff}[X(q)]}~~~~~~,~~~~~
  w[X_s(q)] = \frac{1}{Z}~.
\end{equation}
Hence, a nearly uniform distortion of the gluon distribution
($a\sim0$) is highly suppressed\footnote{Such essentially
  $q$-independent $\eta(q)$ occurs when one selects configurations
  with a high multiplicity above $Q_{s,A}$, i.e.\ in the regime where
  the anomalous dimension is far from its DGLAP
  limit~\cite{Dumitru:2017cwt}. However, here we are interested in
  configurations with a higher than average multiplicity of gluons
  above $Q_{gs,A}$ where $\gamma_A(q)\simeq1$.} since $V_\mathrm{eff}$
is proportional to a power of $Q^2$. For $a>1$, on the other hand, the
penalty action is small but such deviations from the average gluon
distribution do not increase the multiplicity above $Q_{gs,A}$
significantly, and therefore they do not dominate the configuration
bias.  Therefore, from now on we take $a=1$ for illustration. Note the
weak dependence of $V_\mathrm{eff}[X(q)]$ on the lower limit $Q_{s,A}$
of the distortion of the gluon distribution for $a\le1$. Thus, adding
gluons with transverse momenta all the way down to $Q_{s,A}$ does not
come at a high price.

Given a suppression factor (relative weight) $p_r$ we can eliminate
one of the parameters of the trial function, for example
\begin{equation} \label{eq:eta0}
\eta_0 \, q_0^2 \simeq \frac{8\pi\log p_r^{-1}}{N_c^2 A_\perp 
  \log\frac{Q^2}{Q_{s,A}^2}}~.
\end{equation}

Next, to determine $N_\mathrm{coll}[\eta]$ we integrate
eq.~(\ref{Eq:LOapp_Xa}) over $k>Q_{gs,A}$ for the trial
function~(\ref{eq:trialX}) with $a=1$:
\begin{eqnarray}
\frac{dN_{pA}}{dy}\Big|_{\overset{\mathrm{high\, }k}{\mathrm{high~mult}}}
&=& \frac{\pi g^2N_c^2  \mu^2_{0,A} A_\perp}{(2\pi)^3}
\frac{Q_{s,p}^2}{Q_{gs,A}^2}
\left[1+ \frac{\eta_0}{2}
  \frac{q_0^2}{Q_{gs,A}^2}  \right]
  \log \left(\frac{Q_{gs,A}^2}{Q_{s,p}^2}\right)
~.   \label{Eq:LO_Xhigh}
\end{eqnarray}
We approximate $\gamma_\mathrm{DGLAP}\simeq1$.
For a proton target the multiplicity above
$k=Q_{gs,A}$, on average over configurations, is
\begin{eqnarray}
\left<\frac{dN_{pp}}{dy}\right>_{\mathrm{high\, }k}
&=& \frac{\pi g^2N_c^2  \mu^2_{0,p} A_\perp}{(2\pi)^3}
\frac{Q_{s,p}^2}{Q_{gs,A}^2}
  \log \left(\frac{Q_{gs,A}^2}{Q_{s,p}^2}\right) ~.   \label{Eq:LO_p_Xhigh}
\end{eqnarray}
Hence,
\begin{equation}
N_\mathrm{coll}\Big|_\mathrm{high~mult}  =
\frac{\mu^2_{0,A}}{\mu^2_{0,p}}
\left[1+ \frac{\eta_0 q_0^2}{2Q_{gs,A}^2} \right]~.
\end{equation}
Finally, to obtain ${\cal Q}_{pA}(k)$ we divide the gluon spectrum for
a heavy-ion target from eq.~(\ref{Eq:LOapp_Xa}), for the specific
$X(q)$ written in eq.~(\ref{eq:trialX}), by $N_\mathrm{coll}
\big|_\mathrm{high~mult}$ and by the average gluon spectrum for a
proton target:
\begin{equation}
  {\cal Q}_{pA}(k) = \frac{1+\eta_0\left(\frac{q_0^2}{k^2}\right)}
  {1+ \frac{\eta_0}{2}  \frac{q_0^2}{Q_{gs,A}^2}}
  \frac{(k^2)^{\gamma_p(k)-\gamma_A(k)}}
       {Q_{s,A}^{2-2\gamma_A(k)} Q_{s,p}^{2\gamma_p(k)-2}}  =
\frac{1+\eta_0\left(\frac{q_0^2}{k^2}\right)}
  {1+ \frac{\eta_0}{2}  \frac{q_0^2}{Q_{gs,A}^2}}
\, R_{pA}(k) ~.
\end{equation}
In the second step we used the result for $R_{pA}(k)$ written in
eq.~(\ref{eq:RpA_mb}) above. For $k\sim Q_{gs,A}$ we have ${\cal
  Q}_{pA}(k) \simeq1$ since both anomalous dimensions are close to 1,
just like for minimum bias collisions discussed in
eq.~(\ref{eq:RpA_mb}). In fact, ${\cal Q}_{pA}(k)$ as we define it
asymptotically drops below $R_{pA}$ but only slightly, since from
eq.~(\ref{eq:eta0}) it follows that $\eta_0 q_0^2/Q^2_{gs,A}\ll 1$. In
any case, the most interesting regime to us is $k<Q_{gs}$.

The key point is that by selecting
events with a {\em slightly} modified spectrum at high transverse
momentum, $k\sim Q_{gs,A}$, we study the modification at
lower transverse momentum well below $Q_{gs,A}$. In this regime ${\cal
  Q}_{pA}(k) - R_{pA}(k)$ increases with decreasing $k$. With $\eta_0$ from
eq.~(\ref{eq:eta0}),
\begin{eqnarray}
  {\cal Q}_{pA}(k) &\simeq& \left(1+
    \frac{8\pi\log p_r^{-1}}{N_c^2 A_\perp
      k^2\log\frac{Q^2}{Q^2_{s,A}}} \right)
\, R_{pA}(k)~.  \label{eq:QpA_RpA}
\end{eqnarray}
To simplify this expression we have assumed that $\eta_0
q_0^2/Q^2_{gs,A} \ll 1$.  Note that
\begin{eqnarray}
    \frac{8\pi\log p_r^{-1}}{N_c^2 A_\perp
      k^2\log\frac{Q^2}{Q^2_{s,A}}}
\, R_{pA}(k)
\end{eqnarray}
can also be interpreted as the {\em difference} of ${\cal Q}_{pA}(k)$,
$k$ in the range $Q_{s,A}\ll k\ll Q_{gs,A}$, for two classes of events
with relative probability $p_r$. Also, we find that ${\cal Q}_{pA}(k)
- R_{pA}(k)$ increases with decreasing area of the
projectile\footnote{Compare to the numerical analysis of the scaling
  with $A_\perp k^2$ in sec.~\ref{sec:JIMWLKnum} below.}. Away from
midrapidity and towards the hemisphere of the nucleus ${\cal
  Q}_{pA}(k) - R_{pA}(k)$ should increase, and vice versa.

Thus, we have shown in this section that it is possible, in principle,
that a bias on configurations with slightly enhanced gluon density in
the target at high transverse momentum, and correspondingly slightly
higher $N_\mathrm{coll}$, exhibits the features seen in the
data. Namely, that ${\cal Q}_{pA}(k)\simeq1$ at $k\sim Q_{gs,A}$ where
the anomalous dimension is close to its DGLAP limit, while
${\cal Q}_{pA}(k) - R_{pA}(k)$ then increases with decreasing $k$.

\subsection{$\left<\mathrm{tr} \,  W^\dagger (x) W(y) \right>$ to all
  orders in $(gA^+)^4$}

For completeness, in this section we provide the resummed expression
for $\left<\mathrm{tr} \, W^\dagger (x) W(y) \right>$, to all orders
in $A^+$. We did not require the all-order expression for our
considerations above because we restricted to transverse momenta in
the linear regime above the saturation scale.

We first compute explicitly the contribution to $\left< \mathrm{tr}\,
W^\dagger(\vp{x}) W(\vp{y}) \right>$ at fourth order in
$gA^+$:
\begin{eqnarray}
  & &  \frac{1}{2} g^8 N_c^2 (N_c^2-1)
  \int {\cal D}X(\vp q) \, e^{-V_\mathrm{eff}[X(\vp q)]}
\left(
  \int \dd{s}\, \frac{\mu^2(s)}{s^4}
  \left(1-e^{i\vp s\cdot\vp r}\right)
    \frac{X(\vp{s})}{X_s(s)}\right)^2 ~,
\end{eqnarray}
where $\vp r = \vp y - \vp x$.  This is recognized as one half the
square of the contribution at order $(A^+)^2$ written in
eq.~(\ref{eq:O(A+^2)}), except for the overall factor of $(N_c^2-1)$,
a Fourier transform w.r.t.\ $\vp r$, and an integration over
transverse impact parameter space.  This is a consequence of the
ordering in $x^-$ which generalizes to all orders in $A^+$ so that
\begin{eqnarray}
&& \left< \frac{1}{N_c^2-1}\mathrm{tr}\,
W^\dagger(\vp{x}) W(\vp{y}) \right>=
\int {\cal D}X(\vp q) \, e^{-V_\mathrm{eff}[X(\vp q)]}
\, \exp\left(-g^4N_c
  \int \dd{s}\, \frac{\mu^2(s)}{s^4}
  \left(1-e^{i\vp s\cdot\vp r}\right)
    \frac{X(\vp{s})}{X_s(s)}\right)~.  \label{eq:WxWy_all_orders1}
\end{eqnarray}
With $X_s(s)$ from eq.~(\ref{eq:Xs}) we can also write this as
\begin{eqnarray}
&& \left< \frac{1}{N_c^2-1}\mathrm{tr}\,
W^\dagger(\vp{x}) W(\vp{y}) \right>=
\int {\cal D}X(\vp q) \, e^{-V_\mathrm{eff}[X(\vp q)]}
\, \exp\left(-\frac{2}{N_c A_\perp}
  \int \dd{s}  \left(1-e^{i\vp s\cdot\vp r}\right)
    X(\vp{s})\right)~.  \label{eq:WxWy_all_orders2}
\end{eqnarray}
For the trial functions~(\ref{eq:trialX}) we obtain
\begin{eqnarray}
&& \frac{1}{N_c^2-1} \mathrm{tr}\,
W^\dagger(\vp{x}) W(\vp{y}) \Big|_{X(s)=\mathrm{eq.\,}\protect(\ref{eq:trialX})} \sim
\exp\left(- \frac{1}{4} \frac{C_A}{C_F} \frac{1}{2-2\gamma}
\left( r^2 Q_{s,A}^2\right)^\gamma
- \frac{1}{8} \frac{C_A}{C_F} \frac{\eta_0}{a+\gamma-1}  Q_{s,A}^2r^2
\frac{q_0^{2a}}{Q_{s,A}^{2a}}
\right)
\end{eqnarray}
if $r^2 Q^2\ll1$; and
\begin{eqnarray}
&& \frac{1}{N_c^2-1} \mathrm{tr}\,
W^\dagger(\vp{x}) W(\vp{y}) \Big|_{X(s)=\mathrm{eq.\,}\protect(\ref{eq:trialX})} \sim
\exp\left(- \frac{1}{4} \frac{C_A}{C_F} \frac{1}{2-2\gamma}
\left( r^2 Q_{s,A}^2\right)^\gamma
- \frac{1}{8} \frac{C_A}{C_F} \frac{\eta_0}{a+\gamma-1}  q_0^{2a} 
Q_{s,A}^{2\gamma} (r^2)^{\gamma+a}
\right)~.     \label{eq:trWW}
\end{eqnarray}
if $r^2 Q^2\gg1$.  Recall that $Q^2\gg Q_{s,A}^2$ denotes the maximum
transverse momentum for the distortion of the gluon distribution due
to the bias. To arrive at the expressions above we have assumed that
$\gamma$ is constant, that $a+\gamma>1$, and that $r$ is much less
than any non-perturbative infrared cutoff such as the radius of the
proton. Also, throughout the manuscript $Q_s^2$ denotes the saturation
scale for a fundamental dipole which is why the factor
$C_A/C_F\simeq2$ appears. The factor $\eta_0 q_0^{2a}$ can be
expressed in terms of the suppression probability $p_r$ as explained
above. For a nearly fixed $r$, say $r\sim 1/Q_{s,A}$, the correction
in the exponent of eq.~(\ref{eq:trWW}) effectively corresponds to a
shift of the saturation scale. However, this interpretation does not
apply in general since the correction due to the bias involves a
different power of $r^2$ than the unbiased average dipole correlator.

\section{Numerical results for the functional distribution of the
  correlator of adjoint Wilson lines from JIMWLK evolution} 
\label{sec:JIMWLKnum}

In this section we show results of numerical Monte-Carlo simulations
of $\mathrm{tr}\, W^\dagger(\vp q) W(-\vp q)$. We generate random
color charge configurations of the target on a large
lattice\footnote{We used 2048 sites per transverse spatial dimension
  and discretized the longitudinal $x^-$ axis in 100 slices. As a
  cross check we have compared to a $1024^2\times100$ lattice. We
  impose periodic boundary conditions, and global color charge
  neutrality for every configuration. For further details of our
  implementation we refer to ref.~\cite{Dumitru:2014vka}.}  according
to the action of the McLerran-Venugopalan model~\cite{MV} from which
we determine the covariant gauge field $A^+$. These are then evolved
to rapidity $Y>0$ by solving the leading order B-JIMWLK
renormalization group equations~\cite{balitsky,jimwlk} at {\em fixed
  coupling}. The saturation scale $Q_s(Y)$ is determined implicitly
from the forward scattering amplitude of a fundamental dipole: ${\cal
  N}_Y(r=\sqrt{2}/Q_s) = 1-1/\sqrt{e}$. Note that ${\cal N}_Y(r)$ is
averaged over all field configurations. 

We employ a simple reweighting procedure, taking $b_i=1$ for a subset
of 5\% of configurations with the most gluons at high
$p_T>Q_\mathrm{gs}$, and $b_i=0$ for the rest. We do not impose any
constraints such as eq.~(\ref{eq:trialX}) on the spectral shape of the
gluon distribution in this sample, neither below nor above
$Q_\mathrm{gs}$.

For a heavy ion versus a proton target we set the ratio of initial
saturation scales by default to $Q^2_{s,A}(Y=0)/Q^2_{s,p}(Y=0) =
\mu^2_A / \mu^2_p = 6$. This corresponds to the thickness of the
nucleus relative to a nucleon. As already mentioned above we do not
attempt to include Glauber model fluctuations of $N_\mathrm{coll}$ but
focus, instead, on the distribution of functions $\mathrm{tr}\,
W^\dagger(\vp q) W(-\vp q)$ over the MV or JIMWLK ensembles at fixed
$\mu^2_A$. However, for comparison we shall also show results for
$\mu^2_A / \mu^2_p = 2$ and 1.

The correlator of adjoint Wilson lines includes an integral over the transverse overlap
$A_\perp$ of projectile and target since
\begin{equation}
  \mathrm{tr} \,
  W^\dagger (\vp{k}) W(\vp{k}) = \int_{A_\perp} {\mathrm{d}^2 \vp b}
  \int {\mathrm{d}^2 \vp r} \,
  e^{-i\vp{k}\cdot\vp{r}} \, \mathrm{tr}\, W^\dagger\left(\vp{b}+\frac{\vp{r}}{2}\right)
  \, W\left(\vp{b}-\frac{\vp{r}}{2}\right)~.  \label{eq:WkWk_conv}
\end{equation}
The numerical computations of $\mathrm{tr} \, W^\dagger (\vp{k})
W(\vp{k})$ shown below are performed on a large lattice and we need to
explicitly restrict the integration over $d^2\vp{b}$ to the area of a
proton. To do so, in the numerical computations presented in this
section we take
\begin{eqnarray}
  W^G_{ab}(\vp{k}) &=&
  \int d^2\vp{x} \,e^{-i\vp{k}\cdot \vp{x}} \,
  e^{-x^2/2B} \, W_{ab}(\vp{x})~, \\
  \mathrm{tr}\, W^\dagger (\vp{k}) W(\vp{k}) &\to& 
  \mathrm{tr}\, W^{G\dagger}(\vp{k}) \,
  W^G(\vp{k})~.
\end{eqnarray}
We shall focus on transverse momenta $k\gg 1/\surd{B}$ and so the
dominant dipole size $r$ in eq.~(\ref{eq:WkWk_conv}) is set by $\sim
1/k$ and not by $\surd{B}$. Rather, $B<\infty$ restricts the endpoints
of the dipole to within a patch of area $\sim 2\pi B$ around $\vp
b=0$. Our default value is $\sqrt{B}\, Q_{s,p}(Y=0) =
\sqrt{2/3}$. (In physical units this translates to $\sqrt{B}\approx
0.33$~fm and $2\pi B\approx 0.67$~fm$^2$ if $Q_{s,p}(Y=0)\approx
0.5$~GeV.)

Our goal is to determine numerically the function
\begin{equation}
  {\cal Q}(k) = \frac{\left<\mathrm{tr}\, W^{G\dagger} (\vp{k}) W^G(\vp{k})\Big|_A
    \right>_\mathrm{bias}}
  {N_\mathrm{coll}^\mathrm{bias} \left<\mathrm{tr}\, W^{G\dagger} (\vp{k})
    W^G(\vp{k})\Big|_p\right>}
  ~.  \label{eq:Q_k}
\end{equation}
The Wilson lines in the numerator correspond to the field of a heavy
ion while those in the denominator correspond to a proton.  The
corresponding field configurations differ by the choice of initial
saturation scales, $Q^2_{s,A}(Y=0) / Q^2_{s,p}(Y=0) = \mu^2_A /
\mu^2_p >1$, as already mentioned above.  Also, in the denominator we
average over all configurations while the average in the numerator is
biased towards ``high multiplicity'' configurations as follows.

To each configuration we associate a ``number of gluons'' in the
linear regime of high transverse momenta,
\begin{equation}
  N_g = Q_{s,p}^2(Y) \int\limits_{Q_\mathrm{gs}(Y)} {\d^2 \vp k}\,
\mathrm{tr}\, W^{G\dagger}(\vp{k}) \, W^G(\vp{k})
  ~.  \label{eq:Ng}
\end{equation}
This is similar to the integral of eq.~(\ref{Eq:LOapp}) over $\vp k$,
up to a logarithm.  Also, we take $Q_\mathrm{gs}(Y) = \sqrt{3}
\frac{Q^2_{s,A}(Y)}{Q_{s,p}(Y=0)}$. Note that the lower limit of the
integral over $\vp k$ has to increase in proportion to $Q^2_{s,A}(Y)$
in order to avoid the regime where the anomalous dimension differs
significantly from its DGLAP limit.

$\left<\cdots\right>_\mathrm{bias}$ in eq.~(\ref{eq:Q_k}) corresponds
to an average over the subset of configurations with the highest gluon
multiplicity (for example, the 5\% percentile). Also, we define the
``number of binary collisions'', used in eq.~(\ref{eq:Q_k}), for this
subset of configurations via
\begin{equation} \label{eq:Ncoll_MC}
  N_\mathrm{coll}^\mathrm{bias} = \frac{\left< N_g\Big|_A\right>_\mathrm{bias}}
  {\left< N_g\Big|_p\right>}~.
\end{equation}
We emphasize that this quantity is still of order $A^{1/3}$,
proportional to the thickness of the nucleus, because the gluon
distributions for all configurations from the MV/JIMWLK ensembles are
proportional to $A^{1/3}$.  In the absence of the multiplicity bias
$N_\mathrm{coll}\approx Q_{s,A}^2(Y) / Q_{s,p}^2(Y)$ equals the ratio of
the squared saturation momenta\footnote{Fixed coupling evolution
  preserves the proportionality of $Q_{s}^2(Y)$ to $Q_{s}^2(Y=0)$ so
  that $Q_{s,A}^2(Y) / Q_{s,p}^2(Y) = Q_{s,A}^2(Y=0) /
  Q_{s,p}^2(Y=0)$.}. ${\cal Q}(k)$ then simply corresponds to the scaled
ratio of the ``minimum bias'' gluon distributions of a heavy ion
target and a proton.

\begin{figure}[htb]
 \includegraphics[width=8.5cm]{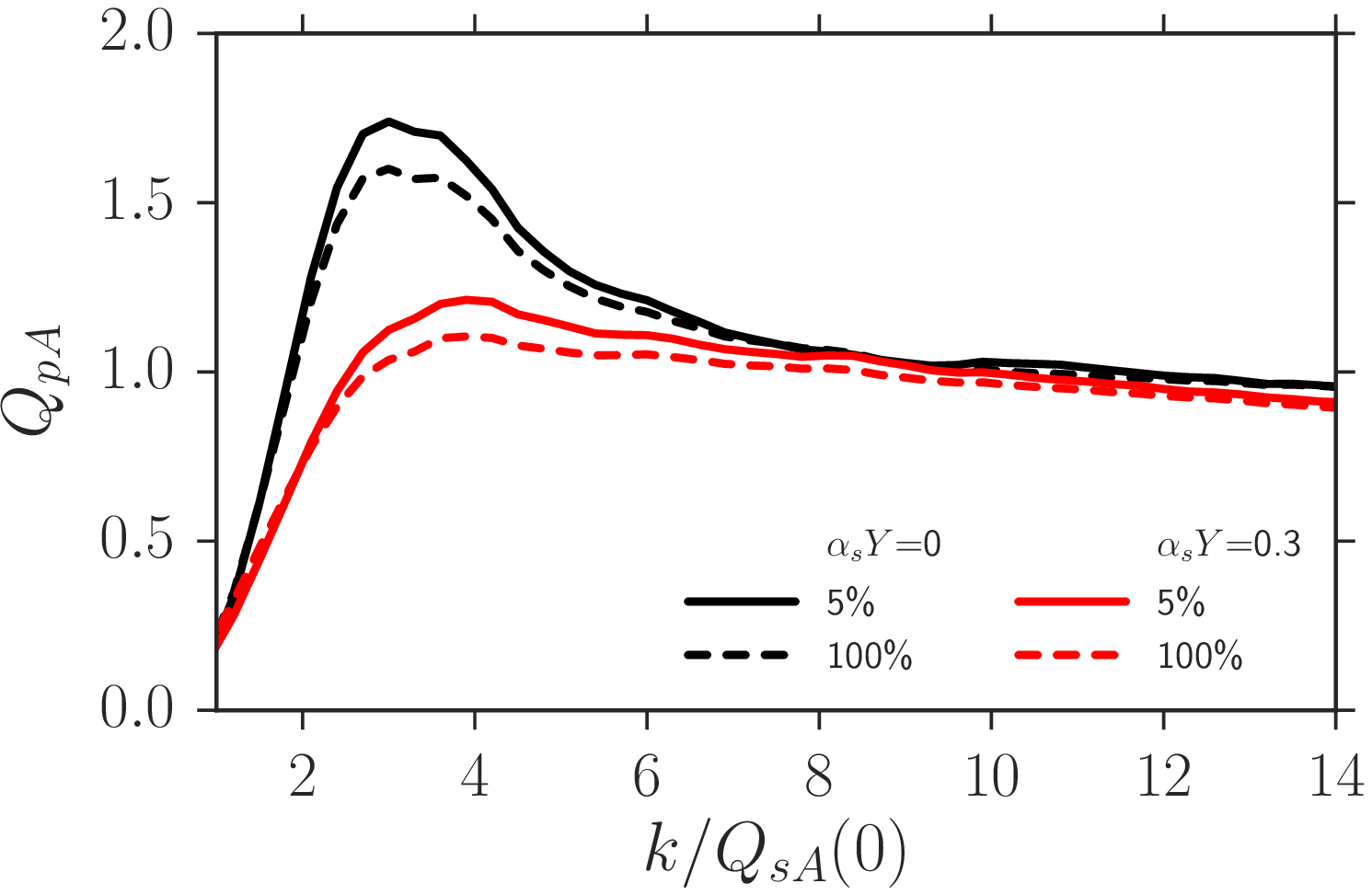}
 \includegraphics[width=8.5cm]{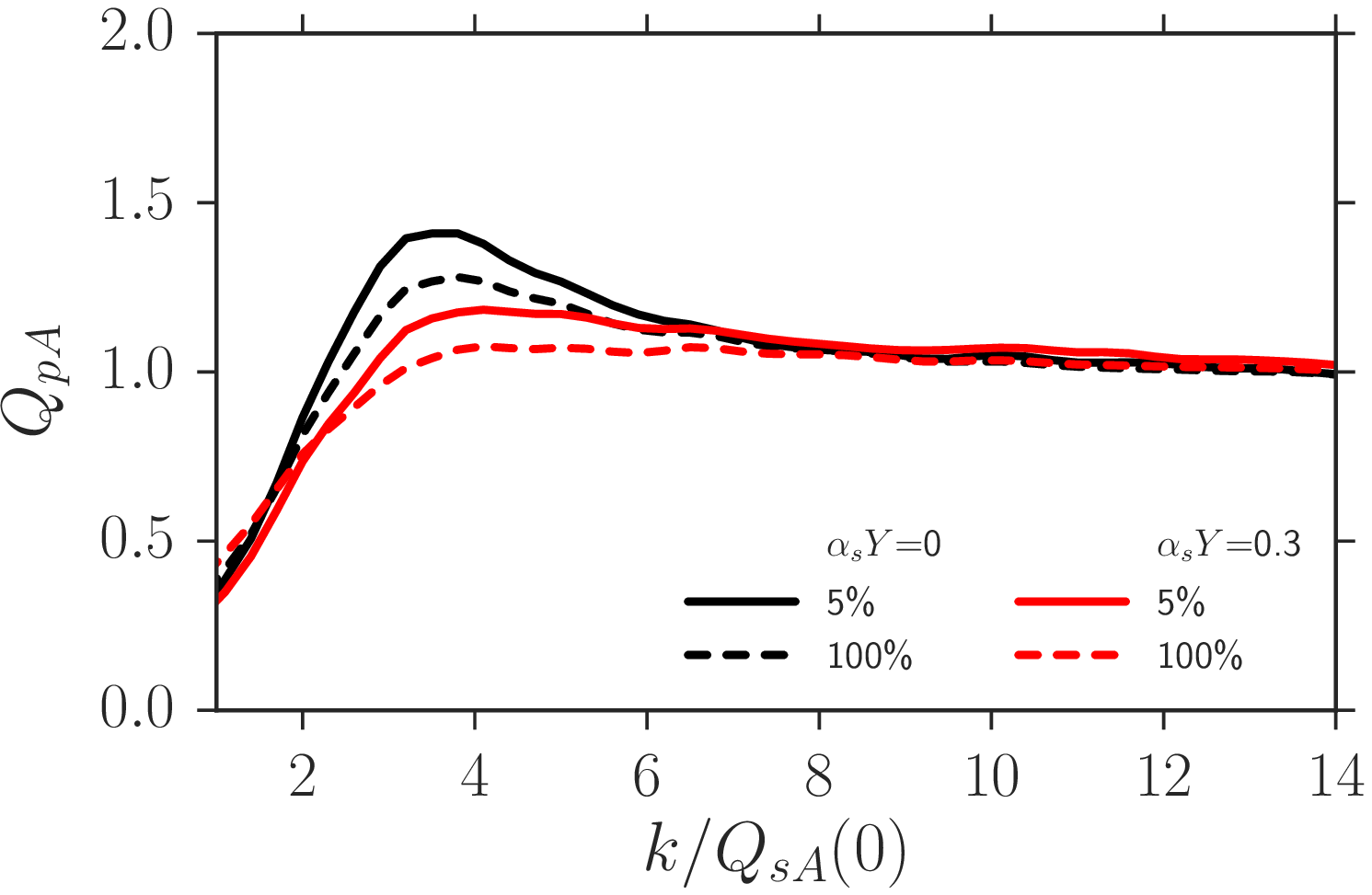}
 \caption{The ratio ${\cal Q}_{pA}(k)$ of $\left<\mathrm{tr}\,
   W^\dagger(k) W(k)\right>$ for a heavy-ion to a proton target
   normalized by $N_\mathrm{coll}$ as defined in
   eqs.~(\protect\ref{eq:Ng},\protect\ref{eq:Ncoll_MC}). This is
   analogous to the ratio of spectra of gluons produced in $pA$
   vs.\ $pp$ collisions divided by $N_\mathrm{coll}$. From top to
   bottom the pairs of curves correspond to the MV model initial
   condition at $Y=0$, and to the JIMWLK ensemble at $\alpha_s Y=0.3$,
   respectively. The dashed curves refer to the unbiased nuclear
   target correlator ($R_{pA}$) while the solid curves correspond to
   subsamples of 5\% of configurations with the most gluons at high
   $k>Q_{gs}(Y)$. For the figure on the left (right) the nucleus is
   assumed to have a thickness of $\mu_A^2/\mu_p^2=6$~(2) nucleons.}
 \label{fig:QpA_JIMWLK}
 \end{figure}
Our result for ${\cal Q}_{pA}(k)$ for a target with a thickness of 6
nucleons is shown in fig.~\ref{fig:QpA_JIMWLK} (left). The magnitude
of the minimum bias ``Cronin peak'' at $Y=0$ is similar to that
obtained from the correlator of fundamental Wilson
lines~\cite{RpA_AAKSW}. With increasing rapidity we observe the
well-known overall suppression of the nuclear modification factor. Our
main interest here is in the effect of the multiplicity bias on this
ratio. Indeed, we observe that a bias on high hard particle multiplicity
drives ${\cal Q}_{pA}(k)$ up. Also, the slope beyond the peak
decreases with increasing evolution rapidity which agrees with the
arguments presented in sec.~\ref{sec:WW_LO}. It is not straightforward
to match the evolution ``time'' $\alpha_s Y$ in these fixed coupling
computations to a collision energy. However, we note that the unbiased
ratio at $\alpha_s Y=0.3$ is quite similar to the measured $R_{pA}(k)$
in minimum-bias p+Pb collisions at 5~TeV\footnote{Clearly, at that evolution
  ``time'' the $R_{pA}$ ratio has not yet fully converged to the fixed
  point of the small-$x$ RG, and its initial shape has not been wiped
  out entirely.}. At that rapidity then, the enhancement of ${\cal Q}_{pA}(k)$
for the 5\% percentile relative to $R_{pA}(k)$ is also not too
different from the ALICE data shown in fig.~\ref{fig:QpA_ALICE}. Thus,
we find that it is important to account for the multiplicity bias on
the small-$x$ gluon field ensemble average.

Fig.~\ref{fig:QpA_JIMWLK} (right) shows ${\cal Q}_{pA}(k)$ for a
target with a thickness of only 2 nucleons. At the initial rapidity
this generates a significantly smaller ``Cronin peak'' than the 6
nucleon target, as expected. On the other hand, JIMWLK evolution leads
to a rather weak dependence on the thickness of the target: the dashed
lines corresponding to $\alpha_s Y=0.3$ in the two panels of
fig.~\ref{fig:QpA_JIMWLK} do not differ by much\footnote{However, the
  transverse momentum scale $Q_{sA}(Y=0)$ in the two panels of
  fig.~\ref{fig:QpA_JIMWLK} does differ by a factor of $\sqrt{3}$
  since it is proportional to $\mu_A$.}. In fact, selecting
the 5\% ``most central'' configurations enhances the peak at
$\alpha_s Y=0.3$ at least as much as increasing the thickness from 2
to 6 nucleons. This illustrates again that at high energy the
centrality selection not only selects events with more target nucleons
but that it also biases significantly the average over small-$x$ gluon
field configurations.

\begin{figure}[htb]
 \includegraphics[width=9.5cm]{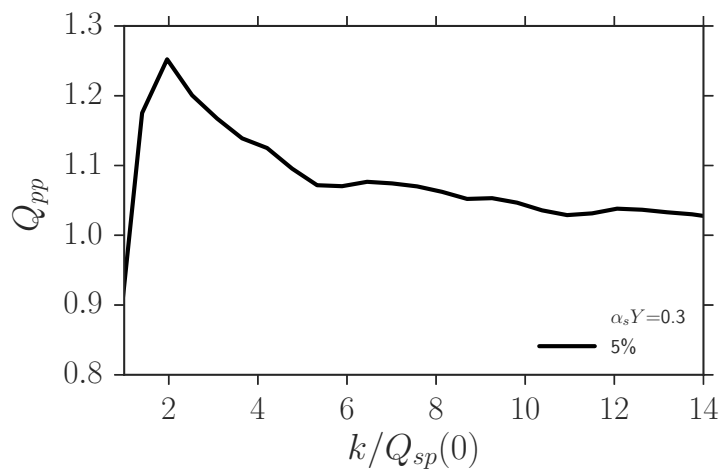}
 \caption{The ratio ${\cal Q}_{pp}(k)$ of $\left<\mathrm{tr}\,
   W^\dagger(k) W(k)\right>$ for central to minimum-bias $pp$ collisions
   normalized by $N_\mathrm{coll}$ as defined in
   eqs.~(\protect\ref{eq:Ng},\protect\ref{eq:Ncoll_MC}). The curve
   corresponds to the JIMWLK ensemble at $\alpha_s Y=0.3$.}
 \label{fig:Qpp}
 \end{figure}
The bias due to selection of a high hard particle multiplicity
modifies the observed gluon distribution even in $pp$
collisions\footnote{Experimentally, ``central'' $pp$ collisions could
  be selected via the transverse energy deposited in a particular
  window in rapidity, analogous to the analysis of $pA$ collisions
  performed by the ATLAS collaboration~\cite{Aad:2016zif}.}. This is
shown in fig.~\ref{fig:Qpp} which corresponds to
$\mu^2_A/\mu^2_p=1$. Hence, the unbiased $R_{pA}(k)=1$ at any
rapidity. However, at $\alpha_s Y=0.3$, when we select configurations
with a higher than average gluon multiplicity at $k>Q_{gs}(Y)$ that
generates an enhancement of ${\cal Q}_{pp}(k)$ relative to the unbiased
ensemble, and a ``Cronin peak'' at $k/Q_{s,p}(0) \simeq 2$. In
practice, this peak occurs at rather low transverse momentum and may
be difficult to observe. However, beyond the peak we clearly observe
${\cal Q}_{pp}(k)-1 >0$ and that, again, it falls off approximately
like $1/k^{2\nu}$ with $\nu=0.6\pm0.1$.

So far we have identified the area $A_\perp$ over which we integrate
the Fourier transform of $\left<\mathrm{tr}\, W^\dagger(x)
W(y)\right>$ with the area of the projectile proton, and used
$\sqrt{B} = \sqrt{2/3}/Q_{s,p}(Y=0)\approx 0.33$~fm, $2\pi B\approx
0.67$~fm$^2$. We now explore the dependence of the biased nuclear
modification factor ${\cal Q}_{pA}(k)$ on the area $A_\perp = 2\pi
B$. We expect that the ensemble of $\mathrm{tr}\, W^\dagger(k) W(k)$
probes larger deviations from the average as the area decreases.

\begin{figure}[htb]
 \includegraphics[width=9.5cm]{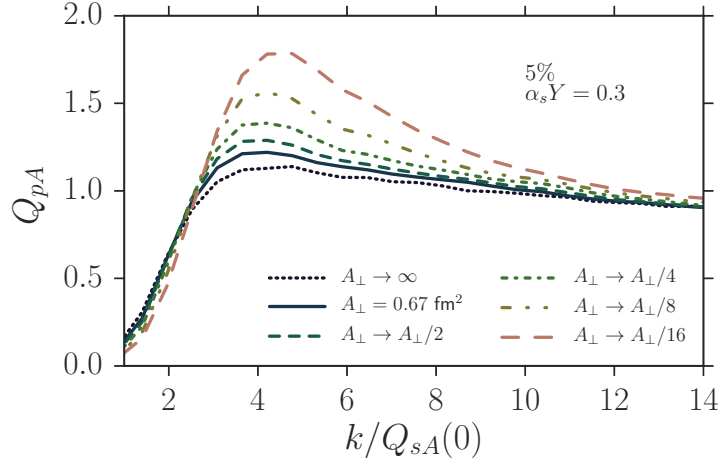}
 \caption{The ratio ${\cal Q}_{pA}(k)$ of $\left<\mathrm{tr}\,
   W^\dagger(k) W(k)\right>$ at $\alpha_sY=0.3$ for the 0-5\%
   centrality class. The target has a thickness of $\mu_A^2/\mu_p^2 =
   6$ nucleons. From bottom to top the curves correspond to an area
   $A_\perp=2\pi B\to \infty$, to the default $2\pi B =
   4\pi/(3Q_{s,p}^2(Y=0)) \approx 0.67$~fm$^2$, and to the default
   rescaled by 1/2, 1/4, 1/8, and 1/16, respectively.}
 \label{fig:QpA_JIMWLK_small}
\end{figure}
In fig.~\ref{fig:QpA_JIMWLK_small} we show ${\cal Q}_{pA}(k)$ for the
5\% centrality percentile for different areas. Taking $B\to\infty$
gives ${\cal Q}_{pA}(k)$ close to the minimum bias nuclear
modification factor since the average over an infinite transverse area
converges toward the average over many independent configurations,
i.e.\ to no bias. For decreasing area one observes, indeed, a stronger
bias due to our ``centrality'' selection, and a more pronounced Cronin
peak.  Ref~\cite{Frankfurt:2003td} suggested that events in which a
very hard dijet ($p_T\sim100$~GeV) is produced may correspond to
configurations of the proton where its small-$x$ gluons occupy a
smaller transverse area. Alternatively, one may tag on the presence of
a $Z$-boson~\cite{atlas-Z}.  If a high-$Q^2$ trigger indeed selects
more compact configurations then it would be very interesting to
analyze ${\cal Q}_{pA}(k)$ for such event samples.

\begin{figure}[htb]
 \includegraphics[width=9.5cm]{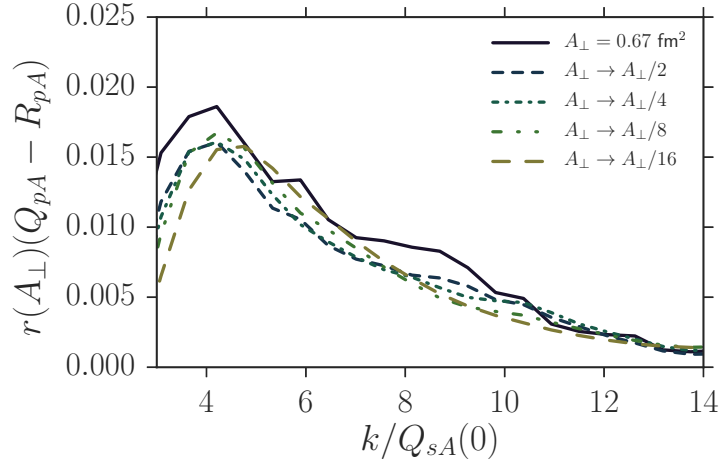}
 \caption{${\cal Q}_{pA}(k)$ for the 0-5\%
   centrality class minus the minimum bias $R_{pA}(k)$, both at rapidity
   $\alpha_sY=0.3$. The target has a thickness of $\mu_A^2/\mu_p^2 =
   6$ nucleons. We rescale the vertical axis by a factor of $r(A_\perp)\sim
   {A_\perp}^\nu$ with $\nu\approx0.6\pm0.1$.}
 \label{fig:QpA_scale_Aperp}
\end{figure}
To exhibit the scaling of ${\cal Q}_{pA}(k)-R_{pA}(k)$ with the
transverse area more clearly in
fig.~\ref{fig:QpA_scale_Aperp} we multiply this quantity by
\be
r(A_\perp) = \left(\frac{A_\perp}{A_\perp^\mathrm{def}}\right)^\nu~.
\ee
Here, $A_\perp^\mathrm{def}=2\pi B\approx0.67$~fm$^2$ represents our
default. For $\nu=0.6\pm0.1$ we observe approximate scaling in that
the curves corresponding to different areas are mapped onto each
other. Thus, for $A_\perp$ not very much smaller than our default
proton projectile the difference of ${\cal Q}_{pA}(k)$ and $R_{pA}(k)$
scales approximately like $\sim 1/{A_\perp}^{0.6}$. Furthermore, beyond
the peak ${\cal Q}_{pA}(k)-R_{pA}(k)$ falls off like $\sim
(k^2)^{-0.6}$. In all,
\be
{\cal Q}_{pA}(k)-R_{pA}(k) \sim \left(A_\perp k^2\right)^{-0.6\pm0.1}~.
\ee
%


\section{Summary and Outlook}
The nuclear modification factor $R_{pA}(p_\perp)$ in
proton-nucleus collisions provides insight into the gluon distribution
of a dense target. It is expected to exhibit leading twist shadowing
for transverse momenta below the so-called extended geometric scaling
scale $Q_{gs}(Y)$. This is due to the fact that the small-$x$ gluon
distribution acquires an anomalous dimension $\gamma_s <
\gamma_\mathrm{DGLAP}$. Furthermore, one expects a stronger
suppression of $R_{pA}(p_\perp)$ with increasing thickness of the
target.

One may also study the nuclear modification factor in a subclass of
$pp$ and $pA$ events, for example in ``central'' collisions as defined
by a suitable centrality selection.  In $pA$ collisions this would
select events where the projectile proton suffers an inelastic
interaction with a greater than average number of target
nucleons. However, such a selection of events also introduces a {\em
  bias} on the ensemble of gluon distribution functions of the target
over which one averages.

In this paper we have provided first analytical and numerical studies
of the biased nuclear (or proton) modification factor ${\cal
  Q}_{pA}(p_T)$ obtained from a reweighted JIMWLK ensemble at small
$x$. The modified $p_T$-dependence of ${\cal Q}_{pA}(p_T)$ as compared
to $R_{pA}(p_T)$ provides insight into the ensemble of gluon
distributions of the small-$x$ fields.

Using the number of high-$p_\perp$ hard particles as a centrality
selector our main observations are as follows. The biased ensemble
exhibits gluon distributions, relative to the average, which increase
with decreasing transverse momentum. That is, more and more extra
gluons appear as $p_\perp$ decreases, so long as it remains greater
than a few times the saturation scale $Q_{s,A}(Y)$ of the target. This
is due to the fact that the action penalty for an additional gluon
increases with transverse momentum. Consequently, we find that the
biased modification factor ${\cal Q}_{pA}(p_\perp)$ is {\em above} the
minimum-bias $R_{pA}(p_\perp)$ and that it may even redevelop a ``Cronin
peak'', despite evolution to small $x$. This is in qualitative
agreement with data taken by the ATLAS~\cite{Aad:2016zif} and
ALICE~\cite{Adam:2014qja} collaborations in centrality selected $p+$Pb
collisions at 5~TeV. In fact, when we integrate the gluon distribution
of the target over a transverse area typical of a proton then the
magnitude of the peak in ${\cal Q}_{pA}(p_\perp)$ is not very
different from what is seen in the data. We predict the same effect
even for $pp$ collisions.

The action penalty for a distorted gluon distribution also increases
with the transverse area $A_\perp$ over which it is integrated. Hence,
we find a stronger increase of ${\cal Q}_{pA}(p_\perp) -
R_{pA}(p_\perp)$, for the 5\% most central collisions say, when the
projectile is in a particularly compact configuration. More
specifically, our numerical results indicate that ${\cal
  Q}_{pA}(p_\perp) - R_{pA}(p_\perp) \sim 1/{(p_\perp^2 A_\perp)}^{0.6\pm0.1}$.
It may be possible to select experimentally such compact projectile
configurations by using a trigger for a hard
dijet~\cite{Frankfurt:2003td} or a $Z$-boson~\cite{atlas-Z}, and to
then analyze the centrality biased modification factor in this
triggered sample.

In the future, it would be interesting to study the effect of a bias
on {\em several} observables simultaneously. For any given reweighting
functional $b[X(\vp q)]$ the shift of all observables $O[X(\vp q)]$ is
determined uniquely. For instance, it will be interesting to see how
two-particle angular correlations~\cite{Kovner:2012jm} are modified in
``central'' $pA$ collisions in coincidence with the $R_{pA}(p_\perp)
\to {\cal Q}_{pA}(p_\perp)$ shift discussed here.

\section*{Acknowledgements}
V.S.\ thanks Anton Andronic for useful discussions and the
ExtreMe Matter Institute EMMI (GSI Helmholtzzentrum f\"ur
Schwerionenforschung, Darmstadt, Germany) for partial support and
their hospitality.

A.D.\ gratefully acknowledges support by the DOE Office of Nuclear
Physics through Grant No.\ DE-FG02-09ER41620; and from The City
University of New York through the PSC-CUNY Research grant 60262-0048.

\end{document}